\documentstyle[12pt]{article}

\begin{document}

\renewcommand{\thefootnote}{\fnsymbol{footnote}}
\renewcommand{\theequation}{\arabic{section}.\arabic{equation}}

\newcounter{saveeqn}
\newcommand{\alpheqn}%
{ \setcounter{saveeqn}{\value{equation}}%
  \stepcounter{saveeqn}\setcounter{equation}{0}%
  \renewcommand{\theequation}%
  {\mbox{\arabic{section}.\arabic{saveeqn}\alph{equation}}}}
\newcommand{\reseteqn}%
{ \setcounter{equation}{\value{saveeqn}}%
  \renewcommand{\theequation}{\arabic{section}.\arabic{equation}}}

\def\refname{References and Footnotes}

\def\ben{\begin{equation}}
\def\een{\end{equation}}
\def\bea{\begin{eqnarray}}
\def\eea{\end{eqnarray}}
\def\nn{\nonumber}

\def\half{\frac{1}{2}}
\def\D{{\cal D}}
\def\Svac{S_0}
\def\Sone{S_1}
\def\S{{\cal S}}
\def\Rvac{\rho_0}
\def\Rone{\rho_1}
\def\ru{\,{}^Ru}
\def\lu{\,{}^Lu}
\def\rb{\,{}^Rb}
\def\lb{\,{}^Lb}
\def\Tr{{\rm Tr}\,}

\let\barr=\overline

\def\BU{\it
                {}\\
                Department of Physics\\
                Boston University\\
                590 Commonwealth Avenue\\
                Boston, Massachusets\ \ 02215\ \ \ U.{}S.{}A.}


%
\hfuzz=5pt
\tolerance=10000
\begin{titlepage}
\begin{center}
{\Large Entanglement Entropy of Nontrivial States}
\footnote
{This work was supported in part by funds provided by the U.{}S.
Department of Energy (D.{}O.{}E.) under contract \#DE-FG02-91ER40676}

\vspace{.4in}
{Eric Benedict\footnote
{e-mail: eben@budoe.{}bu.{}edu}
and So-Young Pi}
\vspace{.4in}
\BU
\end{center}
\vspace{.8in}
\abstract{
We study the entanglement entropy arising from coherent states and
one--particle states. We show that it is
possible to define a finite
entanglement entropy by subtracting the vacuum entropy from that of the
considered states, when the unobserved region is the same.
}
\vfill
\hbox to \hsize{BU-HEP 95-13\hfill May, 1995}
\end{titlepage}

\setcounter{page}{2}

\begin{flushleft}
Proposed running head: Entanglement Entropy

\vspace{1.1in}
Please send correspondence to

\bigskip\medskip
{\hspace*{0.2in}Eric Benedict\\
 \hspace*{0.2in}Department of Physics\\
 \hspace*{0.2in}Boston University\\
 \hspace*{0.2in}590 Commonwealth Avenue\\
 \hspace*{0.2in}Boston, Massachusets 02215}

\medskip
 {\hspace*{0.2in}e-mail: eben@budoe.bu.edu\\
  \hspace*{0.2in}353-6062}

\end{flushleft}
\vfill
\pagebreak
\setlength{\baselineskip}{24pt}

\section{Introduction}

Recently, quantum corrections to the black hole entropy from matter
fields have been studied extensively [1-11].
One source of the quantum
corrections may be understood as entropy of entanglement, which arises
when the density matrix of a pure quantum field theoretic state is
reduced because the quantum field is not observed in some region of
space. It is hoped that this concept provides a description for
some of the contributions to the black
hole entropy, where the unobserved quantum field lies within the black
hole horizon.

In quantum field theoretic calculations, ultraviolet divergences
appear due to the infinite number of degrees of freedom at short
distances. Such infinities arise in the entanglement entropy because
there is an infinite number of states near the boundary between the
observed and unobserved regions. This implies that there is a conflict
between the entropy defined by the counting of quantum states and the
finite Bekenstein--Hawking thermodynamic entropy of a black hole.

In this paper we examine the possiblity of defining a finite
entanglement entropy of a nontrivial state by subtracting the one
associated with the vacuum.
We study the entanglement entropy associated
with coherent states and one--particle states in a massless scalar
field theory in $(1+1)$-dimensional Minkowski space--time. We find that
the entanglement entropy for the coherent states is the same as that for
the vacuum, a result that can be generalized to a space--time of
arbitrary dimension, with an unobserved region of arbitrary shape.
For a restricted class of one--particle states we calculate the
entanglement entropy explicitly and show that, once the vacuum
expression is subtracted, the remainder is finite. We discuss
possible divergences in the
entanglement entropy for more general states.

We begin by presenting a brief review of the entanglement entropy
associated with the vacuum state, which arises when we trace over the
fields in the negative $x$ region by considering an imaginary boundary
at $x=0$. The Hamiltonian of the system is
\ben
H=\half\int dx\:\Pi^2(x)+\half \int dxdy\:
  \Phi(x)\Omega^2(x,y)\Phi(y)
\label{Ham}
\een
where $\Pi(x)$ is the canonical momentum of $\Phi(x)$ and
$\Omega^2(x,y)=-\nabla^2\delta(x-y)$. In the functional Schr\"odinger
representation, the vacuum wave functional has the form
\ben
\langle\phi|0\rangle_M =
\Psi_0[\phi]=\det{}^{1\over 4}\left({\Omega\over\pi}\right)
  \exp\left\{-\half\int dxdy\: \phi(x)\Omega(x,y)\phi(y)
\right\}\label{SchrVac}
\een
where $\phi(x)$ is a $c$-number field at a fixed time (the label $M$
indicates that this is the Minkowski vacuum). By constructing
the pure state density matrix and tracing over the field in the negative
$x$ region, we obtain a reduced density matrix,
\bea
\Rvac(\phi^1_+,\phi^2_+)&=&\int\D\phi_-
  \Psi_0(\phi^1_+,\phi_-)\Psi_0^*(\phi^2_+,\phi_-)\nn\\
&=&\left[\frac{\det \Omega}{\det \Omega_{--}}\right]^{\half} e^{
    -\half\int(\phi^1_+A_{++}\phi^1_+ +
    \phi^2_+A_{++}\phi^2_+ + 2 \phi^1_+B_{++}\phi^2_+
   )
 }\label{RhoZero}
\eea
where $A_{++}$ and $B_{++}$ are functions of the kernel entering
in Eq. (\ref{SchrVac}),
\ben
A_{++}=\Omega_{++}+B_{++}\qquad;\qquad B_{++}=-\half
    \Omega_{+-}\Omega^{-1}_{--}\Omega_{-+}\:.\label{Mats}
\een
[Throughout we use a self--evident functional/matrix notation, with
$\phi_-\equiv\phi(x\!<\!0)$, $\phi_+\equiv\phi(x\!>\!0)$,
$\Omega_{+-}\equiv \Omega(x\!>\!0,y\!<\!0)$ {\it etc.},
  and $\int\phi A\phi\equiv
\int\int dxdy\:\phi(x) A(x,y) \phi(y)$.]
In Eq. (\ref{Mats}) $\Omega^{-1}_{--}$ is the inverse of the restricted kernel
$\Omega_{--}$. It has been shown first by Bombelli {\it et~al.}
\cite{BKLS}, that $\Rvac$ may
be diagonalized by solving the eigenvalue problem
\ben
\int_0^{\infty}
dz\:\Lambda_{++}(x,z)\psi(z)=\lambda\psi(x)\:.\label{EigProb}
\een
where
\ben
\Lambda_{++}(x,z)\equiv-\int_{-\infty}^0dy\:
  [\Omega^{-1}]_{+-}(x,y)\Omega_{-+}(y,z)\:.
\een
In this expression $[\Omega^{-1}]_{+-}$ is the inverse of the full kernel with
argument restricted.
The eigenfunctions and eigenvalues of $\Lambda_{++}$ are
\alpheqn\bea
\psi(x)&\propto&\exp(ik\ln x)\label{EigFunc}\\
\lambda(k)&=&{1\over\sinh^2\pi k}\:.\label{EigVal}
\eea\reseteqn
To calculate the entropy
we must discretize the spectrum. We adopt
the procedure used in Ref \cite{CalWil}: we introduce an infrared
cutoff $L$ and an ultraviolet cutoff $\epsilon$
, {\it i.{}e.} $\epsilon\leq x\leq L$,
and demand that $\psi(x)$
vanish at $x=L$ and $x=\epsilon$,
\alpheqn\bea
\psi(x)&=&\sin(k_n\ln x/\epsilon)\label{EigFuncDisc}\\
k_n&=&{\pi n\over\ln(L/\epsilon)}\quad,\qquad n\mbox{ an integer.}
  \label{EigValDisc}
\eea\reseteqn
The vacuum entanglement entropy can be approximated as an integral,
\ben
\Svac\quad=\quad\sum_n\S_0(k_n)
 \quad\approx\quad \frac{2}{\pi}\ln\frac{L}{\epsilon}
  \int_{0}^{\infty}d\omega\: \S_0(\omega)\label{VacEnt}
\een
where $\S_0(\omega)$ is the contribution of the eigenmode
$\omega\equiv|k|$,
\ben
\S_0(\omega)=-\ln(1-\mu)-{\mu\over 1-\mu}\ln\mu\:\quad;\quad
 \mu=e^{-2\pi\omega}\:.\label{OmegaVacEnt}
\een
This contribution is finite as $\omega\rightarrow\infty$, but
diverges as $\omega\rightarrow 0$. The divergence is integrable,
however, and the integral in Eq. (\ref{VacEnt}) is finite. As can be seen
from Eq. (\ref{VacEnt})
$\Svac$ is infinite as $\epsilon\rightarrow 0$ due
to the infinite density of states near the boundary at $x=0$.

\section{Entanglement Entropy of a Coherent State}
\setcounter{equation}{0}

We consider a coherent state
$|\Psi_\alpha\rangle$, which is an eigenstate of the
annihilation operator,
\alpheqn\ben
a(x)={1\over\sqrt 2}\left[\int_{-\infty}^{\infty} dy\: \Omega^\half(x,y)\Phi(y)
  +i\int_{-\infty}^{\infty} dy\: \Omega^{-\half}(x,y)\Pi(y)\right]\label{Ann}
\een
\ben
\langle\phi|a(x)|\Psi_\alpha\rangle=\alpha(x)\Psi_\alpha[\phi]
  \label{CohState}
\een\reseteqn
where $\alpha(x)$ is in general
complex. The solution to Eq. (\ref{CohState}) is
\ben
\Psi_\alpha[\phi]=N \exp\left[-\half\int\int\phi \Omega\phi+
  \sqrt{2}\int\int\alpha \Omega^{1/2}\phi\right]\:.\label{PsiSol}
\een
For our calculation it is useful to write $\Psi_\alpha$ in terms of
the real and imaginary parts of its (functional) eigenvalue
$\alpha\equiv\alpha_R+i\alpha_I$:
\ben
\Psi_\alpha[\phi]=Ne^{i\int\barr\pi\phi}e^{-\half\int\int
  (\phi-\barr\phi)\Omega(\phi-\barr\phi)}\label{PsiComp}
\een
where
\alpheqn\bea
\barr\phi(x)&=&\sqrt{2}\int_{-\infty}^{\infty}dy\:
   \alpha_R(y) \Omega^{-\half}(y,x)
  \label{PhiZero}\\
\barr\pi(x)&=&\sqrt{2}\int_{-\infty}^{\infty}dy\:
   \alpha_I(y) \Omega^{\half}(y,x)
  \label{PiZero}
\eea\reseteqn
and a factor $\exp(-\half\int\barr\phi\Omega\barr\phi)$ has been absorbed
into the normalization.
By tracing over $\phi_-$ we obtain the reduced density matrix for
this state:
\bea
\rho_\alpha(\phi^1_+,\phi^2_+)&=&\int\D\phi_-
  \Psi_\alpha(\phi^1_+,\phi_-)\Psi^*_\alpha(\phi^2_+,\phi_-)\nn\\
&=&e^{i\int\barr\pi_+(\phi^1_+-\phi^2_+)}\rho_0
  (\phi^1_+-\barr\phi_+,\phi^2_+-\barr\phi_+)\label{RhoAlpha}
\eea
where $\rho_0$ is the reduced density matrix for the vacuum state
given in Eq. (\ref{RhoZero}), with the argument $\phi_+^{1,2}$ translated by
$\barr\phi_+$. One can show that the phase appearing in $\rho_\alpha$
disappears in the functional integration involved in the entropy
calculation. It is straightforward to see that the eigenvalues of
$\rho_\alpha$ are the same as those of $\rho_0$, and one concludes
that the
entanglement entropy arising from the coherent state is given by the
vacuum state entanglement entropy:
\bea
S_\alpha&=&-\Tr\rho_\alpha\ln\rho_\alpha\nn\\
&=&-\Tr\rho_0\ln\rho_0\nn\\
&=&S_0\:.\label{SalphSzero}
\eea

We can understand this result with the following argument. The
relation between $\rho_\alpha$ and $\rho_0$ given in Eq. (\ref{RhoAlpha})
reflects the fact that the coherent state labelled by $\alpha$ is
related to the vacuum state by a unitary transformation. This becomes
clear if we rewrite Eq. (\ref{PsiComp}) in terms of the unitary
operator constructed from $\Pi$ and $\Phi$,
\ben
\Psi_\alpha[\phi]=N\langle\phi|
 e^{i\int\barr\pi\Phi}e^{i\int\barr\phi\Pi}|0\rangle_M\:.\label{OpCohState}
\een
Moreover, when we factor the basis state into
$|\phi\rangle=|\phi_+\rangle\otimes|\phi_-\rangle$, the unitary
operator factors into two pieces, one acting on $|\phi_+\rangle$
alone and
the other acting on $|\phi_-\rangle$ alone, due to the commutation
relation $[\Pi_{\pm},\Phi_{\mp}]=0$:
\ben
e^{i\int\barr\pi\Phi}e^{i\int\barr\phi\Pi}
=\left(e^{i\int\barr\pi_+\Phi_+}e^{i\int\barr\phi_+\Pi_+}\right)
\left(e^{i\int\barr\pi_-\Phi_-}e^{i\int\barr\phi_-\Pi_-}\right)\:.
\label{SepFact}
\een
Eqs. (\ref{OpCohState}) and (\ref{SepFact}) then lead to Eq.
(\ref{SalphSzero}).

Our result is rather surprising, since the entropy we calculate is
determined by the counting of quantum states, and there is no reason
to expect that the entanglement entropy arising from the two
different pure states should be the same. This result may be easily
generalized to a spacetime of arbitrary dimension and to an
unobserved region of any shape.

\section{Entanglement Entropy of One--Particle States}
\setcounter{equation}{0}

In this section we study the structure of the divergences of the
entanglement entropy arising from
one--particle states. Calculation of
the entanglement entropy associated with one--particle states is in
general extremely complicated due to the difficulty in diagonalizing
the reduced density matrix.
We find that the calculation is much
simpler if we use an alternative quantization of Minkowski space,
employing the Rindler space mode functions \cite{Unruh}.
Therefore, we shall first describe this formalism.

\subsection{Rindler Space Description of the Minkowski Vacuum}

We introduce the familar two--wedge Rindler coordinates $\xi$ and
$\eta$, which are related to the Minkowski coordinates $x$ and $t$
through the relations
\alpheqn\bea
x&=&\pm a^{-1}e^{a\xi}\cosh(a\eta)\label{RindX}\\
t&=&\pm a^{-1}e^{a\xi}\sinh(a\eta)\label{RindT}
\eea\reseteqn
where the parameter $a$ is a positive constant, and the sign is taken
to be positive and negative in the right--hand and left--hand
wedges, respectively. The Rindler
coordinates take all real values, $-\infty<\xi,\eta<\infty$, and
cover the two quadrants
of Minkowski space given by $|x|>|t|$. Due to the fact that
the Rindler metric is conformal
to all of Minkowski space and the massless Klein--Gordon equation is
conformally invariant in $(1+1)$--dimensions, there exist mode
solutions of the form
\alpheqn\bea
\ru_k&=&\cases{
  {1\over\sqrt{4\pi\omega}}e^{i(k\xi-\omega\eta)}&in right wedge\cr
  0&in left wedge\cr}\label{RUK}\\
\nn\\
\lu_k&=&\cases{
  0&in right wedge\cr
  {1\over\sqrt{4\pi\omega}}e^{i(k\xi+\omega\eta)}&in left wedge\cr}
  \label{LUK}
\eea\reseteqn
where $\omega=|k|$.
These mode functions can be analytically continued to the region
$|x|<|t|$, and
together, $\lu_k$ and $\ru_k$ are complete in all of Minkowski space
\cite{Boulware}.
The field operator $\Phi$, therefore, may be expanded in terms of
these solutions, resulting in an alternative Fock space;
\ben
\Phi=\int dk\:\left(
\lb_k\lu_k+\rb_k\ru_k+\mbox{h.c.}
\right)\label{PhiRind}
\een
The operators $\rb_k$ and $\lb_k$ annihilate the Rindler vacuum,
$|0\rangle_{\rm Rind}\equiv|0\rangle_R\otimes|0\rangle_L$,
\ben
\lb_k|0\rangle_{\rm Rind}
  \quad=\quad\rb_k|0\rangle_{\rm Rind}\quad=\quad 0\:.
  \label{BDef}
\een
With this alternative formulation one can characterize the states of
the quantum field either as
Minkowski particle states according to the usual quantization, in
which the field operator is expanded as
\ben
\Phi=\int{dp\over\sqrt{4\pi\omega}}\:\biggl[
a_p e^{i(px-\omega t)}+a_p^{\dag}e^{-i(px-\omega t)}
\biggr]\label{MinState}
\een
with $\omega=|p|$,
or as Rindler particle states according to Eq. (\ref{PhiRind}).

We note that at a fixed time $t=\eta=0$ the
Rindler mode functions in Eq. (\ref{RUK}, \ref{LUK}), with $a=1$,
are precisely the eigenfunctions in (\ref{EigFunc}) that diagonalize the
vacuum reduced density matrix given in Eq. (\ref{RhoZero}).
When we write the instantaneous field
configuration as
\ben
\phi(\xi(x))=\theta(x)\int_{-\infty}^{\infty}
  \frac{dk}{2\pi}\phi_R(k)e^{ik\xi(x)}+
\theta(-x)\int_{-\infty}^{\infty}
  \frac{dk}{2\pi}\phi_L(k)e^{ik\xi(x)}\label{PhiInst}
\een
the Minkowski vacuum wave functional [Eq. (\ref{SchrVac})] has the
following form in the basis $|\phi_L\phi_R\rangle$ \cite{Freese}:
\bea
\Psi_0(\phi_R,\phi_L)&=&\langle\phi_L\phi_R|0\rangle_M\nn\\
 &\propto&\exp\left\{
-\half\int_{-\infty}^{\infty}\frac{dk}{2\pi}\:\left[
    k\coth(\pi k/a)\biggl(|\phi_R|^2+|\phi_L|^2\biggr)
    -\frac{2k\,{\rm Re}\,\phi_R\phi^*_L}{\sinh(\pi k/a)}
  \right]
\right\}\:.\nn\\
 \label{RSchrVac}
\eea
By constructing the density matrix for this state
and integrating over the field
configuration for $x<0$, we obtain the reduced density matrix
\bea
\rho_0(\phi^1_R,\phi^2_R)&=&
  \int\D\phi_L\D\phi_L^*\:\Psi_0(\phi_R^1,\phi_L)\Psi_0^*(\phi_R^2,\phi_L)\nn\\
&\propto&\exp\left\{
-\half\int_{-\infty}^{\infty}\frac{dk}{2\pi}\:\left[
    k\coth(2\pi k/a)\biggl(|\phi^1_R|^2+|\phi^2_R|^2\biggr)
    -\frac{2k\,{\rm Re}\,\phi_R\phi^*_L}{\sinh(2\pi k/a)}
  \right]
\right\}\:.\nn\\
&&\label{RVacDens}
\eea
This is precisely the density matrix of Eq. (\ref{RhoZero})
when it is diagonalized using Eq. (\ref{EigFunc}, \ref{EigVal}).
The momentum variable $k$, which we shall call the Rindler
momentum, labels the eigenmode with eigenvalue $\lambda(k)$.
Eq. (\ref{RVacDens}) has the form of a thermal density matrix at a
temperature $T_R=a/2\pi$.

It is clear from the above discussion that the entanglement
entropy for the Minkowski space
geometry we are considering may be
analyzed in the language of Rindler space.
In order to discuss the one--particle
Minkowski states we introduce the elegant formulation given in
\cite{Unruh}, where one constructs the
Minkowski positive frequency mode functions
\alpheqn\bea
f_1&=&{1\over\sqrt{2\sinh\pi\omega/a}}\biggl[
   e^{\pi\omega/2a}\ru_k+e^{-\pi\omega/2a}\lu^*_{-k}
  \biggr]\label{Fone}\\
f_2&=&{1\over\sqrt{2\sinh\pi\omega/a}}\biggl[
   e^{-\pi\omega/2a}\ru^*_{-k}+e^{\pi\omega/2a}\lu_k
  \biggr]\label{Ftwo}
\eea\reseteqn
and expands the field as
\ben
\Phi=\int dk\:\biggl(
d_k^{(1)}f_1+d_k^{(2)}f_2+\hbox{h.c.}
\biggr)\:.\label{DExpand}
\een
The operators $d_k^{(1,2)}$ are then given by
\alpheqn\bea
d_k^{(1)}&=&{1\over\sqrt{2\sinh\pi\omega/a}}\biggl[
   e^{\pi\omega/2a}\rb_k-e^{-\pi\omega/2a}\lb^{\dag}_{-k}
  \biggr]\label{DOneDef}\\
d_k^{(2)}&=&{1\over\sqrt{2\sinh\pi\omega/a}}\biggl[
   e^{\pi\omega/2a}\lb_k-e^{-\pi\omega/2a}\rb^{\dag}_{-k}
  \biggr]\:.\label{DTwoDef}
\eea\reseteqn
and they annihilate the Minkowski vacuum,
\ben
d_k^{(1,2)}|0\rangle_M=0\:.\label{DAnn}
\een
This can be explicitly verified in the functional Schr\"odinger
representation, where the vacuum wave functional is given by Eq.
(\ref{RSchrVac}), and ${}^{L,R}b_k$ and ${}^{L,R}b^{\dag}_k$ have
the following form:
\alpheqn\bea
{}^{L,R}b_k&=&\frac{1}{\sqrt 2}\left[
  \sqrt{\omega}\phi_{L,R}(k)+\frac{1}{\sqrt\omega}
    \frac{\delta\ \ \ }{\delta\phi_{L,R}^*(k)}
\right]\\\label{BkSchr}
{}^{L,R}b^{\dag}_k&=&\frac{1}{\sqrt 2}\left[
  \sqrt{\omega}\phi_{L,R}^*(k)-\frac{1}{\sqrt\omega}
    \frac{\delta\ \ \ }{\delta\phi_{L,R}(k)}
\right]\:.\label{BkDagSchr}
\eea\reseteqn
Eq. (\ref{DAnn}) implies that $d^{(1,2)}_k$
can be expressed as a superposition of
the conventional Minkowski space annihilation operators $a_p$,
\ben
d_k^{(1,2)}=\int_{-\infty}^\infty
  {dp\over 2\pi}\:D^{(1,2)}(k,p)a_p\:.\label{DExp}
\een
We find the expansion coefficients to be
\ben
D^{(n)}(k,p)=\left[{2k\over |p|}\sinh(\pi k/a)\right]^{\half}
   \:\Gamma(k/ia)
   |p|^{ik/a}\:
   \Biggl\{\matrix{\theta(kp)\:,&n=1\cr
                  \theta(-kp)\:,&n=2\cr}
\label{DCoeff}
\een
where $\Gamma(x)$ and $\theta(x)$ are the gamma
and step functions.

To calculate the entanglement entropy for a one--particle state,
we find it convenient to use an
alternative representation of the Minkowski vacuum, constructed in Ref
\cite{Unruh}:
from Eqs. (\ref{DOneDef}), (\ref{DTwoDef}), and
(\ref{DAnn}), we have
\bea
|0\rangle_M&=&\prod_{k,-k}N_k\sum_n
   e^{-\pi n\omega/a}|n_k\rangle_R\otimes|n_{-k}\rangle_L\nn\\
&\equiv&\prod_{k,-k}|0\rangle^k_M\:.\label{MinAsRin}
\eea
The normalization is $N_k=(1-e^{-2\pi\omega/a})^{1/2}$.
The product $\prod_{k,-k}$ is taken over a complete set
of Rindler modes, and $|n_k\rangle_{R,L}$ denotes right and left Rindler
states with $n$ particles of Rindler momentum $k$.
By tracing over the degrees of freedom in the
left region, we obtain the reduced density operator, $\hat\Rvac$;
since it is diagonalized by the Rindler mode functions, it may be
written as a product of density operators, one for each Rindler mode:
\ben
\hat\Rvac\equiv\prod_{k,-k}\hat\Rvac^k\label{RhoHatZero}
\een
where the contribution from each mode is given by
\bea
\hat\Rvac^k&=&\sum_m {}_L\langle m_{-k}|0\rangle^k_M
  {}^{\ k}_M\langle 0|m_{-k}\rangle_L\nn\\
&=&N_k^2\sum_m e^{-2\pi m\omega/a}|m\rangle_R{}_R\langle m|\:.\label{RhoHatK}
\eea
This is an alternative form of the density matrix
given in Eq. (\ref{RVacDens}). The entanglement entropy can be
calculated simply, and, upon discretizing the spectrum by demanding
that $\ru_k$ vanish at $\xi(L)$ and $\xi(\epsilon)$,
we find that it
has the same form as Eq. (\ref{VacEnt}). The advantage of using this
formalism is that it automatically gives the diagonal form of the
reduced density matrix for the one--particle state that we are going
to consider.

\subsection{Entanglement Entropy of a One--Particle State}

We shall now compute the entanglement entropy of a one--particle
Minkowski state with a definite Rindler momentum $k$,
by exciting the positive--frequency mode $f_1$. This is a
particular superposition of Minkowski momentum eigenstates:
\ben
d_k^{(1)\dag}|0\rangle_M=\int_{-\infty}^{\infty}{dp\over2\pi}\:
  D^{(1)*}(k,p)a^{\dag}_p|0\rangle_M\:.\label{RindEigInt}
\een
Using the notation of Eq. (\ref{MinAsRin}),
we write this state as
\ben
d^{(1)\dag}_k|0\rangle_M=|1\rangle_M^k\prod_{\ell\neq k}
  |0\rangle_M^{\ell}\label{RindEig}
\een
where
\bea
|1\rangle_M^k&=&{\cal N}\sum_ne^{-\pi n\omega/a}\biggl\{
\sqrt{n+1}|(n+1)_k\rangle_R\otimes|n_{-k}\rangle_L\nn\\
&&\hbox{\hskip 1.5truein}
   -e^{-\pi\omega/a}\sqrt{n}|n_k\rangle_R\otimes|(n-1)_{-k}\rangle_L
\biggr\}\nn\\
\nn\\
&=&{\cal N}\:2\sinh(\pi\omega/a)\sum_n e^{-\pi n\omega/a}\sqrt{n}|n_k\rangle_R
   \otimes|(n-1)_{-k}\rangle_L\label{OneK}
\eea
The normalization factor is singular:
\ben
{\cal N}^2={}_M\langle 0|d_k^{(1)}d_k^{(1)\dag}|0\rangle_M=2\pi\delta(0)\:.
  \label{SingNorm}
\een
The reduced density matrix, normalized to have unit trace,
is then
\ben
\hat\rho(k)=
{1\over 2\pi\delta(0)}\Tr_L\biggl[
   d_k^{(1)\dag}|0\rangle_M{}_M\langle 0|d_k^{(1)}
  \biggr]=
\hat\Rone^k\prod_{\ell\neq k}\hat\Rvac^{\ell}\label{RhoRind}
\een
where $\Tr_L$ represents a trace over $\{\bigotimes_{k,-k}|n_k\rangle_L\}$,
$\hat\Rvac^{\ell}$ is given by Eq. (\ref{RhoHatK}), and
\ben
\hat\Rone^k=4\sinh^2(\pi\omega/a)\sum_n ne^{-2\pi n\omega/a}
  |n_k\rangle_R{}_R\langle n_k|\:.\label{RhoOne}
\een
The density operator is again diagonal, and the entanglement entropy
is readily calculated:
\ben
S=\S_1(k)+\sum_{\ell\neq k}\S_0(\ell)\:.\label{SRind}
\een
(The summation reflects the fact that the spectrum must be
discretized for this expression to be well--defined.)
When we subtract the entanglement entropy of the vacuum state, $S_0$,
from $S$, $\Delta S\equiv S-S_0$ reduces to
\bea
\Delta S&=&\S_1(k)-\S_0(k)\nn\\
&=&-\ln(1-\mu)-{\mu\over 1-\mu}\ln\mu-
  {(1-\mu)^2\over\mu}\sum_m (m\ln m)\mu^m \label{DeltSExp}
\eea
where $\mu=e^{-2\pi\omega/a}$ as before.
We have evaluated the last term in this expression numerically:
as shown in Fig 1, $\Delta S$ is
finite for all values of $\omega$. It can also be evaluated
analytically in the limits $\omega\rightarrow 0$ and
$\omega\rightarrow\infty$ \cite{OmegaLimit}:
\alpheqn\bea
\lim_{\omega\rightarrow 0}\Delta S&=&\gamma_E\approx 0.5772
   \label{LimZero}\\
\lim_{\omega\rightarrow\infty}\Delta S&=&0 \:.\label{LimInf}
\eea\reseteqn
We emphasize that the density of states factor, which made the vacuum
state entanglement entropy [Eq. (\ref{VacEnt})] diverge, does not enter into
$\Delta S$. Thus we have shown that the entanglement entropy for a
particular class of one--particle states can be made finite by
subtracting from it the vacuum entanglement entropy.

\subsection{More General One--Particle States}

We have computed the entanglement entropy associated with the
one--particle state given in Eq. (\ref{RindEigInt}) in closed form using
the fact that the Rindler space formalism leads to a diagonal reduced
density matrix arising from that state. More general
one--particle Minkowski states are created by operators corresponding
to both the $f_1$ and $f_2$ modes
[see Eqs. (\ref{Fone},\ref{Ftwo})]:
\ben
|\psi\rangle=\int_{-\infty}^{\infty}{dk\over\sqrt 2\pi}
  \:\biggl[
\psi_1(k)d_k^{(1)\dag}+\psi_2(k)d_{-k}^{(2)\dag}
\biggr]|0\rangle_M\:.\label{GenState}
\een
Here $\psi_{1,2}(k)$ are smearing functions determining a
superposition of momentum states.
With such states the problem of diagonalizing the reduced density
matrix remains. Already, for the simple choice
$\psi_1(k)=\psi_2(k)=\sqrt\pi\delta(k-\ell)$, {\it i.{}e.}
\ben
|\psi\rangle={1\over\sqrt 2}\biggl(
  d_\ell^{(1)\dag}+d_{-\ell}^{(2)\dag}
\biggr)|0\rangle_M\label{OneTwoState}
\een
the reduced density matrix is not diagonal. However, we have
computed the entanglement entropy numerically: when the vacuum
contribution is subtracted off, the entanglement entropy
$\Delta S\equiv S-S_0$ is found to be finite (see Fig 2). Therefore,
it is reasonable to expect that, for well--behaved smearing functions
that fall off sufficiently fast at large $k$, the entanglement
entropy arising from the state $|\psi\rangle$ is finite once the
vacuum contribution is subtracted.

In the appendix we provide a formal, but concrete, calculation that
supports the expectation
that the entanglement entropy $\Delta S$ of one--particle
states is finite for states defined by well--behaved smearing
functions.

Our calculation presented in this paper suggests that the difference
between the entanglement entropy arising from different states, but
for the same unobserved region, is finite for a given theory of matter
fields. This is complimentary to another finite quantity that has
recently been studied in \cite{Holzhey}, where the excited state was
produced by a moving mirror.

\begin{appendix}
\bigskip\bigskip
\noindent{\Large\bf Appendix}
\bigskip
\renewcommand{\theequation}{\mbox{A.\arabic{equation}}}
\setcounter{equation}{0}
\renewcommand{\alpheqn}%
{ \setcounter{saveeqn}{\value{equation}}%
  \stepcounter{saveeqn}\setcounter{equation}{0}%
  \renewcommand{\theequation}%
  {\mbox{A.\arabic{saveeqn}\alph{equation}}}}
\renewcommand{\reseteqn}%
{ \setcounter{equation}{\value{saveeqn}}%
  \renewcommand{\theequation}{\mbox{A.\arabic{equation}}}}

\noindent
We shall consider a one--particle state defined by
\ben
|\psi\rangle=\int_{-\infty}^{\infty}{dk\over\sqrt{2\pi}}\:\psi(k)
  d_k^{(1)\dag}|0\rangle_M\label{AFDef}
\een
and discuss possible divergences in the entanglement entropy
$S(\psi)$ associated with $|\psi\rangle$. First, using a variational
principle, we show that $S(\psi)$ is bounded
by an entropy defined by
\alpheqn\bea
S_b&=&\Tr \hat\rho_b\ln\hat\rho_b \label{SBDef}\\
\hat\rho_b&=&\int_{-\infty}^{\infty}dk\:|\psi(k)|^2\hat\rho(k)
  \label{RhoBDef}
\eea\reseteqn
where $\hat\rho(k)$ is the reduced density matrix associated with the
state $d_k^{(1)\dag}|0\rangle_M$ given in Eq. (\ref{RindEig}). We can obtain
an explicit expression for $S_b$ due to the fact that $\hat\rho_b$ is
diagonal in the basis of $n$-particle Rindler states. Then we shall
study the divergence structure of $S_b$, and show that
$\Delta S_b\equiv S_b-S_0$ is finite for well--behaved smearing
functions $\psi(k)$. Although the bound $S(\psi)\leq S_b$ (which we
shall prove below) is a formal one, in that both $S_b$ and $S_0$
are infinite, our calculation is interesting since we show that the
difference, $\Delta S_b$, is finite for well--behaved smearing
functions. The finiteness of the difference supports the expectation
that $\Delta S(\psi)\equiv S(\psi)-S_0$ is finite.

We first present a general extremum principle.

\medskip\noindent{\bf Lemma:}
\medskip

\noindent
Let $\hat\rho(q)$ be a
family of density operators labelled by a continuous parameter $q$,
and consider operators of the form
\ben
\hat\rho[h]=\int dq\:h(q)\hat\rho(q)\:.\label{ARhoMuDef}
\een
We guarantee that $\hat\rho[h]$ will be a density operator by
demanding that $h(q)>0$ and $\int dq\:h(q)=1$. Subject to these
constraints we extremize the functional
\ben
F[h]=-\Tr\hat\rho[h]\ln\hat\rho[h]+\lambda\biggl(
  \int dq\:h(q)-1\biggr)\:.\label{AFMuDef}
\een
Upon setting $\delta F[h]/\delta h(q)=0$ we find the extremum condition
\ben
-\Tr\hat\rho(q)\ln\hat\rho[h]=1-\lambda\:.\label{AExtCond}
\een
The right hand side of this expression is independent of $q$, so Eq.
(\ref{AFMuDef}) is extremized by the $h(q)$ that makes the left hand
side constant as well.\bigskip

Next we construct a family of density operators
$\hat\rho_\psi(q)$ in the following way: define a set of states
generalizing Eq. (\ref{AFDef}),
\ben
|\psi,q\rangle=\int_{-\infty}^{\infty}{dk\over\sqrt{2\pi}}\:q^{-ik}\psi(k)
  d_k^{(1)\dag}|0\rangle_M\label{AFqDef}
\een
and compute the reduced density operator
\ben
\hat\rho_\psi(q)=\Tr_L|\psi,q\rangle\langle \psi,q|\:.\label{ARhoFDef}
\een
In the calculations that follow we shall use a scaling relation for
the states $|\psi,q\rangle$ that is manifest in the functional
Schr\"odinger representation described in section (3.1).
Under the transformation $\phi_{R,L}\rightarrow q^{ik}\phi_{R,L}$, the
one--particle state
\ben
\langle\phi_L\phi_R|d_k^{(1)\dag}|0\rangle_M=N_k\left[
\phi_R^*(k)-e^{-\pi\omega/a}\phi_L^*(k)
\right]\Psi_0(\phi_R,\phi_L)\label{ASchrRep}
\een
(where $N_k$ is a normalization constant) becomes
\ben
\langle q^{ik}\phi_L,q^{ik}\phi_R|d_k^{(1)\dag}|0\rangle_M=
 q^{-ik}\langle\phi_L\phi_R|d_k^{(1)\dag}|0\rangle_M\:.\label{AScaleD}
\een
Combining Eqs. (\ref{AScaleD}) and (\ref{AFqDef}),
we find the scaling relation that we seek,
\ben
\langle\phi_L\phi_R|\psi,q\rangle=
 \langle q^{ik}\phi_L,q^{ik}\phi_R|\psi,1\rangle\:.\label{AScaleF}
\een
Furthermore, the functional measures $\D\phi_R\D\phi^*_R$
and $\D\phi_L\D\phi^*_L$ are
invariant under this transformation.

Now we use the lemma:
consider the extremum condition Eq. (\ref{AExtCond}), which we write as
\ben
-\Tr\hat\rho_\psi(q)\ln\left[
\int_0^{\infty}dq'\:h(q')\hat\rho_\psi(q')
\right]=\hbox{constant}\:.\label{AFExtCond}
\een
Using Eq. (\ref{AScaleF}), and the invariance of the functional measure,
we have
\ben
-\Tr\hat\rho_\psi(q)\ln\left[
\int_0^{\infty}dq'\:h(q')\hat\rho_\psi(q')
\right]=
-\Tr\hat\rho_\psi(1)\ln\left[
\int_0^{\infty}dq'\:h(q')\hat\rho_\psi(q'q^{-1})
\right]
\label{AUseScale}
\een
which is $q$-independent if $dq'h(q')=d(qq')h(qq')$. The normalized
$h(q)$ with this property is
\ben
h(q)={1\over 2\pi\delta(0)}\:{1\over q}\label{AMuSoln}
\een
where we have written the singular normalization
$(\int_0^{\infty}dq\:q^{-1})^{-1}$ as $(2\pi\delta(0))^{-1}$.
The corresponding density operator $\hat\rho[h]$ is
\bea
\hat\rho[h]&=&{1\over 2\pi\delta(0)}\int_{-\infty}^{\infty}dk\:
   |\psi(k)|^2\:\Tr_L\biggl[d_k^{(1)\dag}|0\rangle_M
                {}_M\langle 0|d_k^{(1)}\biggr]\nn\\
&=&\int_{-\infty}^{\infty}dk\:|\psi(k)|^2\hat\rho(k)
  \label{ARhoMuSoln}
\eea
where $\hat\rho(k)$ is given by Eq. (\ref{RhoRind}).
This density operator is
our $\hat\rho_b$ in Eq. (\ref{RhoBDef}). It
is diagonal in the basis $\{\bigotimes_{k,-k}|n_k\rangle_R\}$.

We must show that the extreme value is a maximum for $F[h]$, if it
is to provide a bound. We take the second functional derivative,
evaluated at the extremum,
\ben
{\delta^2 F[h]\over\delta h(q)\delta h(q')}=
 -\Tr{\hat\rho(q)\hat\rho(q')\over\hat\rho_b}\:.\label{AScndDeriv}
\een
(We take the trace with respect to the
basis in which $\hat\rho_b$ is diagonal.) The functional Hessian
matrix, Eq. (\ref{AScndDeriv}), is diagonalized by the set of basis functions
$q^{ik-1}$:
\ben
-\int_0^{\infty}dq\:q^{-ik-1}\int_0^{\infty}dq'\:q'^{ik'-1}
  \Tr{\hat\rho(q)\hat\rho(q')\over\hat\rho_b}
  =-2\pi\delta(k-k')\int_0^{\infty}d\xi\:\xi^{ik'-1}
  \Tr{\hat\rho(1)\hat\rho(\xi)\over\hat\rho_b}\:.
\label{AIsDiag}
\een
We clarify this expression by
defining $\hat{\cal O}(k)=\int_0^{\infty}dq\:q^{ik-1}\hat\rho(q)$,
and writing the diagonal elements as
$-\Tr\hat{\cal O}^{\dag}(k)\hat{\cal O}(k)/\hat\rho_b$.
They are seen to be
the negative trace of the ratio of two positive operators, and are therefore
negative. The eigenvalues of the Hessian matrix are thus all
negative, and the extremum is a maximum. We have therefore
established the formal bound $S(\psi)\leq S_b$.

We shall now calculate $S_b$. We first discretize the spectrum by
introducing infrared and ultraviolet cutoffs as given in Eq.
(\ref{EigValDisc}).
This leads to the following expression for $\hat\rho_b$:
\ben
\hat\rho_b=\lim_{\Delta k\rightarrow 0}\sum_n \Delta k|\psi(k_n)|^2
  \hat\rho(k_n)\label{RhoBExpr}
\een
where $\Delta k=\pi/\ln(L/\epsilon)$, $k_n=n\Delta k$, and
$\hat\rho(k_n)$ is given by (using Eqs. (\ref{RhoHatK}) and
(\ref{RhoOne})),
\ben
\hat\rho(k_n)=\lim_{\Delta k\rightarrow 0}{1-\mu_n\over\mu_n}\prod_i
 \left[
  (1-\mu_i)\sum_{m_i=0}^{\infty}m_n \mu_i^{m_i}|m_i\rangle_R
    {}_R\langle m_i|
 \right]\label{RhoJApp}
\een
with $\mu_n=e^{-2\pi|k_n|/a}$. With the expression above for
$\hat\rho_b$, the entropy $S_b$, when the vacuum entanglement entropy
is subtracted, reduces to
\bea
\Delta S_b&\equiv& S_b-S_0\nn\\
&=&-\sum_n {f_n\ln\mu_n\over 1-\mu_n}-\prod_i(1-\mu_i)
  \sum_{m_1,m_2,\ldots=0}^{\infty}\mu_n^{m_n}
  F(f)\ln F(f) \label{DeltSBound}
\eea
where we have defined $f_n\equiv|\psi_n|^2\Delta k$, and
\ben
F(f)\equiv\sum_n {1-\mu_n\over\mu_n}f_n m_n\:.\label{FPsiDef}
\een
This gives an upper bound for $\Delta S(\psi)$:
\ben
\Delta S(\psi)=S(\psi)-S_0\:\leq\Delta S_b\:.
\een
Although $\Delta S_b$ cannot be evaluated exactly for arbitrary
$\psi(k)$, it provides the following information: due to the
normalization of $\hat\rho_b$, which implies the normalization of the
smearing function $\sum_n \Delta k|\psi(k_n)|^2=1$,
the density of states does not appear
in $\Delta S$ as an overall factor. Thus the only possible source of
divergence is the smearing function $\psi(k)$. However, one can
evaluate Eq. (\ref{DeltSBound}) for smearing functions that satisfy
$\psi(k)=0$ for $|k|>k_0$, where $k_0\ll a$. In this limit, $\Delta
S_b$ is well approximated by a multidimensional integral: if we set
$y_n=-m_n\ln\mu_n$, and $dy_n\sim-\ln\mu_n$, then
\bea
\Delta S&\approx& 1-\int_0^{\infty}\cdots\int_0^{\infty}
  dy_1dy_2\ldots\ e^{-\sum_n y_n}\sum_n y_n f_n\ln
    \sum_n y_n f_n\nn\\
&=&1-\|f\|\ln\|f\|-\|f\|(1-\gamma_E)\label{SmallApprox}
\eea
where $\|f\|^2\equiv\sum_n|f_n|^2\leq 1$. For any of the allowed values of
$\|f\|$ this expression is finite and nonzero. It is reasonable to
suppose that $\Delta S_b$ will remain finite for smearing functions
that drop off sufficiently fast at large $k$: this supports the
expectation that, for resonable values of the smearing function, the
difference between the entanglement entropy of the state given by Eq.
(\ref{AFDef}) and that of the vacuum is finite.

\end{appendix}

\pagebreak

\pagebreak
\noindent{\large\bf Figure Captions}

\bigskip\bigskip
\noindent
Fig. 1: $\Delta S$ for the state in Eq. (\ref{RindEigInt})
as a function of $k/a$. The dashed line is $\gamma_E$.

\bigskip
\noindent
Fig. 2: $\Delta S$ for the state in Eq. (\ref{OneTwoState})
as a function of $\ell/a$.


\begin{thebibliography}{99}
\bibitem{OmegaLimit}
We obtain the $\omega\rightarrow 0$ limit by writing Eq.
(\ref{DeltSExp}) as
\ben
\Delta S=-{\mu\over 1-\mu}\ln\mu-{(1-\mu)^2\over\mu}
 \sum_m m\ln [m(1-\mu)]\mu^m\nn
\een
Defining $x=2\pi\omega/a$, so that $\mu=e^{-x}$, and taking the limit
$x\rightarrow 0$,
\ben
\lim_{x\rightarrow 0}\Delta S
  = \lim_{x\rightarrow 0} 1-x^2\sum_m m\ln(xm)e^{-xm}
  = 1-\int_0^{\infty}dy\:y\ln y e^{-y}
  = \gamma_E\nn
\een
\end{thebibliography}
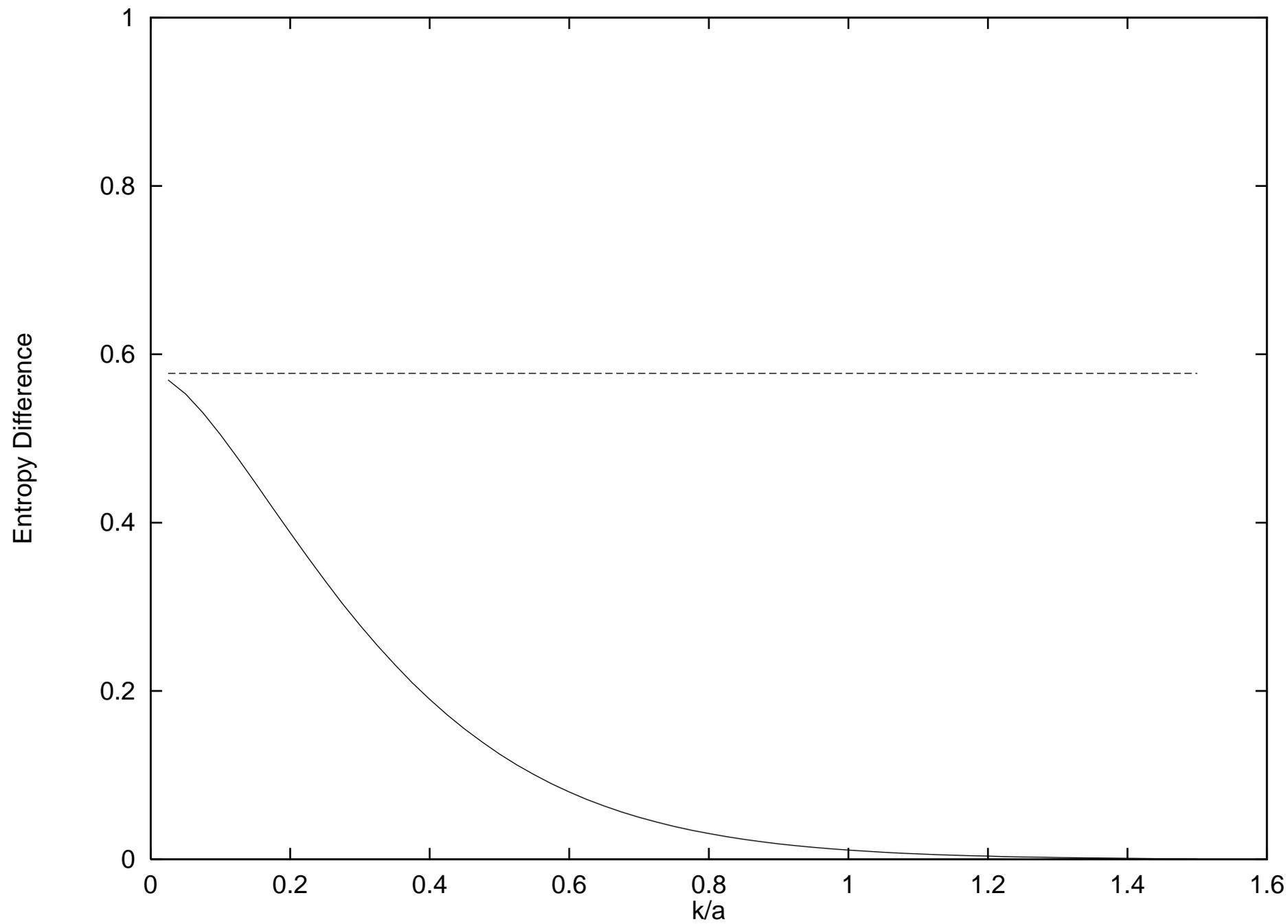
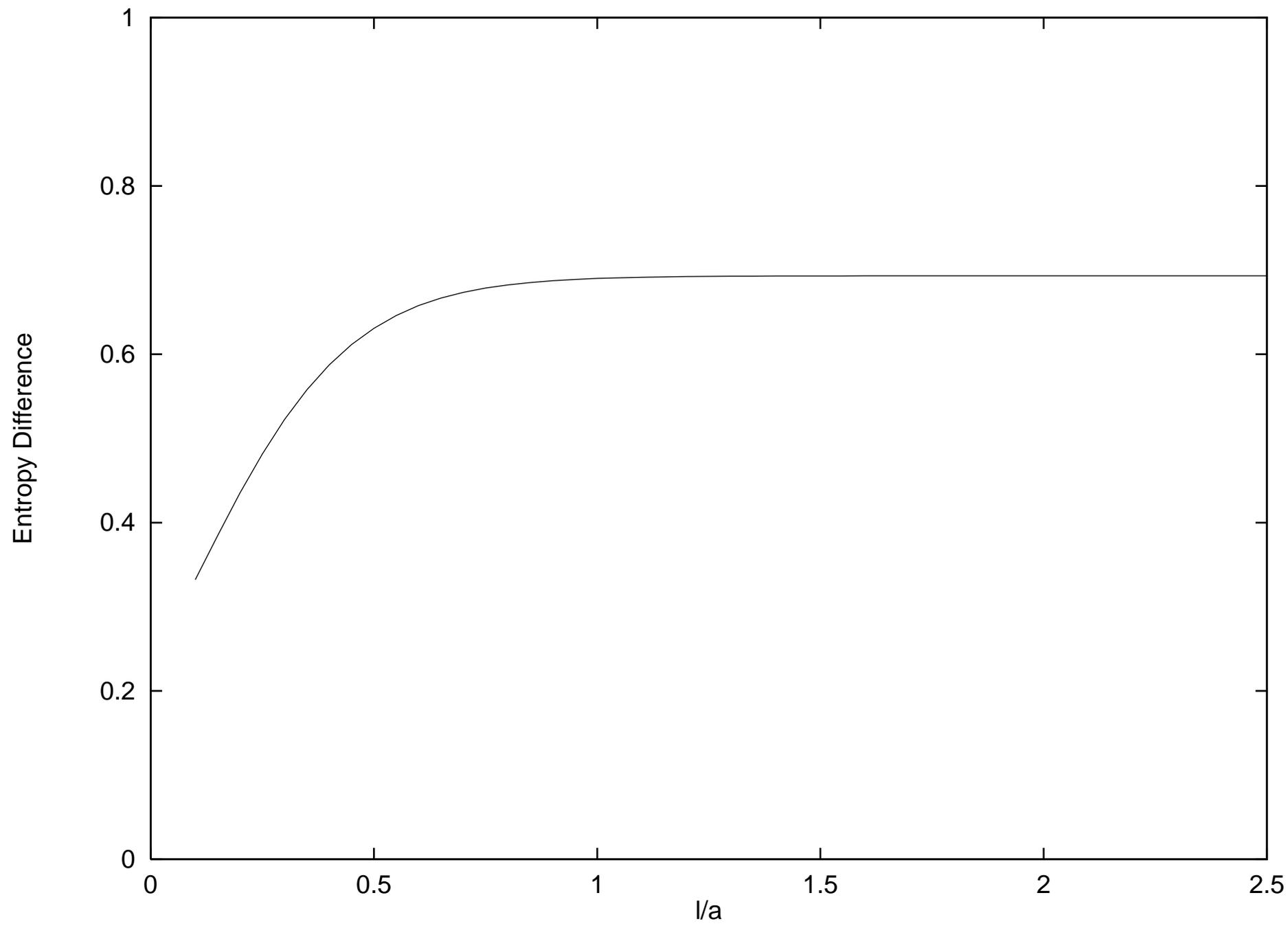
\end{document}